\documentclass[final,3p,times,twocolumn]{elsarticle}
\def\ph2{{\it p}-H$_2$}
\def\od2{{\it o}-D$_2$}

\usepackage{amssymb}

\journal{Results in Physics}

\begin{document}

\begin{frontmatter}



\title{Morphology of dipolar Bose droplets}


\author[inst1]{Massimo Boninsegni}

\affiliation[inst1]{organization={Department of Physics},
            addressline={University of Alberta}, 
            city={Edmonton},
            postcode={T6G 2E1}, 
            state={Alberta},
            country={Canada}}



\begin{abstract}
The ground state of a free standing, self-bound droplet comprising four hundred dipolar Bose particles with aligned dipole moments, with an additional purely repulsive two-body interaction, is investigated by Quantum Monte Carlo simulations.  The focus here is on the evolution of the cluster as the effective range of the repulsive interaction is varied. We identify a ``classical'' regime, in which binding arises exclusively from the dipolar potential energy and the cluster is a quasi-one-dimensional filament, and a ``quantum'' regime of prolate droplets, held together to a significant degree by quantum-mechanical exchanges. The transition between the two regimes occurs abruptly.

\end{abstract}




\end{frontmatter}


\section{Introduction}
\label{intro}
The progress achieved in the stabilization of large  assemblies of cold atoms or molecules featuring permanent electric or magnetic dipoles moments, enables the search of  phases and phenomena not observed in ordinary condensed matter, underlain by the strongly anisotropic character of the interaction  (see, for instance, Ref. \cite{menotti}). 
A chief example is the {\em supersolid} phase of matter \cite{rmp}, whose existence, in different forms, has been predicted in the bulk phase of a  gas of dipolar bosons with aligned dipole moments \cite{spivak,pupillo,Petrov,patterned}. Recently, corroborating experimental evidence for some of these theoretical predictions has been reported \cite{tanzi,fabian,innsbruck}.

Of significant theoretical interest is also the study of  dipolar Bose clusters, i.e., free-standing, self-bound finite assemblies. The experimental and theoretical investigation of highly quantal atomic or molecular clusters, such as $^4$He \cite{skc,grebenev,tang} or H$_2$ \cite{sck,sartakov,prozument}, dates back a few decades, and is motivated by the intriguing physical properties that a finite system displays, which can be markedly different from those of the bulk. As the size of the cluster is increased, its physics approaches that of the bulk, often in a highly non-monotonic way (as in the case of parahydrogen clusters), giving rise to remarkable effects, such as {\em quantum melting} \cite{mezz,mezz2}).

Dipolar clusters (or, droplets) \cite{pfau} offer in principle an even wider, richer playground, owing to the anisotropy, but also the {\em tunability} of the pairwise interaction. In order to see how the latter comes about, one must first note that a purely dipolar Bose system, in which all dipoles moments are aligned along a direction of particle motion, is unstable against collapse \cite{landau}. Stabilization requires the presence of an additional short-range repulsive interaction, physically separate from the dipolar part. Such a  repulsive term has been generally described theoretically as a contact interaction, formally through the so-called scattering length approximation. In the context of cold atoms, the scattering length ($a_s$) is experimentally controllable by means of the Feshbach resonance (see, for instance, Ref. \cite{feshbach}); this renders it possible to adjust the relative strength of the attractive and repulsive parts of the pair potential, exploring different physical limits. 
\\ \indent
On very general grounds, two different, well-defined physical regimes can be identified for a dipolar Bose cluster. In the ``classical'' regime, in which $a_s << a_D$ (the characteristic length of the dipolar interaction), the energetics is dominated by  the potential energy (specifically, the dipolar part thereof), the cluster is strongly self-bound  and takes the form of a  quasi-one-dimensional filament in the direction of dipolar alignment. On the other hand, in the ``quantum'' regime, whose onset is expected as $a_s\gtrsim \gamma \ a_D$, where $\gamma$ is a number of order one,  the attractive part of the dipolar interaction is progressively weakened by the repulsion, and the cluster morphs into a prolate droplet held together to an increasingly significant degree by quantum-mechanical exchanges. As $a_s$ exceeds some critical value, one that increases monotonically with $N$ and saturates to a bulk value, the cluster ceases to be self-bound. Not much theoretical investigation has been carried out on the evolution of the system as $a_s$ is varied, and on whether there exist intermediate regimes displaying intriguing new physics.
\\ \indent
It should be stressed that gaining quantitative understanding of the physics of a single cluster is not only interesting in its own right, especially in light of experiments conducted with individual droplets \cite{pfau}, it is also directly relevant to the understanding of the physics of some of the bulk phases of the system \cite{wachtler,cb,baillie,fb,seesee}.
\\ \indent
This paper describes the results of a theoretical study, based on first principle computer simulations, of a free standing, self-bound dipolar cluster comprising $N=400$ particles, in its ground state. We make use of a theoretical model equivalent to that described above, utilized by us and other authors in previous studies, in which the scattering length approximation is replaced by a purely repulsive, inverse-power-law potential of effective range $\sigma$.
\\ \indent
The main result of this study is that the character of the cluster changes fairly abruptly for a specific value $\sigma^\star\approx 0.28\ a_D$ of the effective diameter of the repulsive interaction, corresponding to a scattering length $a_s \sim 0.15\ a_D$. For  $\sigma<\sigma^\star$ the droplet is a nearly one-dimensional filament whose length grows with $\sigma$, as increasing the hard core diameter increases the equilibrium distance between adjacent particles. Exchanges of identical particles occur infrequently, less and less so as $\sigma\to 0$, and remain fairly localized in character,   typically involving only relatively small groups ($\lesssim 10$) of contiguous particles. As $\sigma\to\sigma^\star$, exchanges become prevalent and the filament takes on a cigar shape, shrinking in length by almost a factor three. On increasing $\sigma$ above $\sigma^\star$, the droplet expands in all directions, radially more than axially, becoming unbound for $\sigma\sim\ 0.5\ a_D$. In the quantum regime, clusters are homogeneously superfluid at low temperature.
\\ \indent 
The remainder of this paper is organized as follows: in section \ref{mm} we describe the microscopic model of the system; in Sec. \ref{me} we briefly describe our methodology; we present and discuss our results in Sec. \ref{else} and finally outline our conclusions in Sec. \ref{conc}.

\section{Model}
\label{mm}
The system is modeled as an ensemble of $N$ identical particles of mass $m$ and spin zero, hence obeying Bose statistics, moving in free space. These particles possess a permanent dipole moment $d$ (either electric or magnetic), pointing in the  $z$-direction, alignment being achieved by means of an external field. 
We are interested in studying {\em free standing} clusters; thus, we do not confine the simulated system by means of an external potential. Rather, our system is enclosed in a three dimensional cubic cell; as commonly done in simulations of finite system, we use for convenience  periodic boundary conditions in all three directions, taking the side of the cell large enough to make boundary conditions irrelevant. 

We take the characteristic length of the dipolar interaction, $a_D \equiv md^2/\hbar^2$ as our unit of length, and $\epsilon \equiv \hbar^2/(ma_D^2)$, as that of energy and temperature.  The quantum-mechanical many-body Hamiltonian in dimensionless form reads as follows:
\begin{eqnarray}\label{u}
\hat H = - \frac{1}{2} \sum_{i}\nabla^2_{i}+\sum_{i<j}U({\bf r}_i,{\bf r}_j)
\end{eqnarray}
where the first (second) sum runs over all particles (pairs of particles), and the pair potential $U({\bf r},{\bf r}') = U_{sr}(|{\bf r}-{\bf r}'|) + U_d({\bf r},{\bf r}')$, $U_{sr}$ being the repulsive part. 
As mentioned above, in most theoretical studies, especially those based on mean-field techniques, the repulsive part of the interaction is modeled by means of the well-known scattering length approximation.
To the extent that such an approximation is valid, any potential that has the same scattering length will yield the same results. In this work, we use for $U_{sr}$ 
the repulsive part of the standard Lennard-Jones potential, i.e.,
\begin{equation}\label{Usr}
U_{sr}(r)=(\sigma/r)^{12}
\end{equation}
which has been used in previous works as well \cite{patterned,cb}, and whose use is convenient in numerical simulations. The parameter $\sigma$ of the potential $U_{sr}$ used here is directly related to the scattering length $a_s$, through
\begin{equation}\label{as2}
{a_s} = b\ \sigma^{1.2}
\end{equation}
with $b\approx 0.71$
(see, for instance, Ref. \cite{flugge}).
$U_{d}$ is the classical dipolar interaction between two aligned dipole moments, namely
\begin{equation}\label{Ud}
U_d({\bf r},{\bf r}')=\frac{1}{|{\bf r}-{\bf r}'|^3}\left(1-\frac{3(z-z')^2}{|{\bf r}-{\bf r}'|^2}\right)
\end{equation}
At zero temperature, there is only one parameter that governs the physics of the system, namely the effective diameter $\sigma$ of the repulsive interaction. In this work, we consider the range $0.18\le \sigma\le 0.49$.

All the results presented here are for a fixed number of particles in the cluster, namely $N=400$. We postpone the discussion of this particular choice to Sec. \ref {conc}. Because, as mentioned above, we are interested in free-standing self-bound clusters, we make use of no external confining potential. Thus, it is always possible that in the course of a long simulation run, one or more particles will ``evaporate'', i.e., break free from the cluster and wander around in the cell. In order to prevent this from happening, in previous simulation studies of small clusters ($N\lesssim 100$) of weakly bound systems such as $^4$He (see, for instance, Ref. \cite{skc}), an artificial confining potential was utilized, whose purpose was exclusively that of keeping the cluster  together. In this study, we have not made use of such a trick, as evaporation does not prove in practice a significant problem (it is essentially unnoticeable for $\sigma \lesssim 0.4$). However, we have adopted the operative criterion of considering a simulation result reliable and representative of the physics of the cluster, if the estimated fraction of evaporated particles is 1\% or less of the total system (i.e., four particles or less).

\section{Methodology}
\label{me}
The computational methodology adopted here is the canonical \cite{mezz,mezz2} continuous-space Worm Algorithm \cite{worm,worm2}. This is a finite temperature ($T$) quantum Monte Carlo (QMC) technique; since we are ultimately interested in ground state properties, we need to extrapolate the results to $T=0$. 
Finite temperature QMC methods have several advantages over ground state ones, for investigating Bose systems, even at $T=0$ (for an extensive discussion  of this subject, see for instance Ref. \cite{rmp95}); in particular, they are unaffected by the bias of existing ground state methods, arising from the use of a trial wave function, as well as from the control of a population of walkers \cite{psb,phasesep}. 

\begin{figure}[h]
\centering
\includegraphics[width=\linewidth]{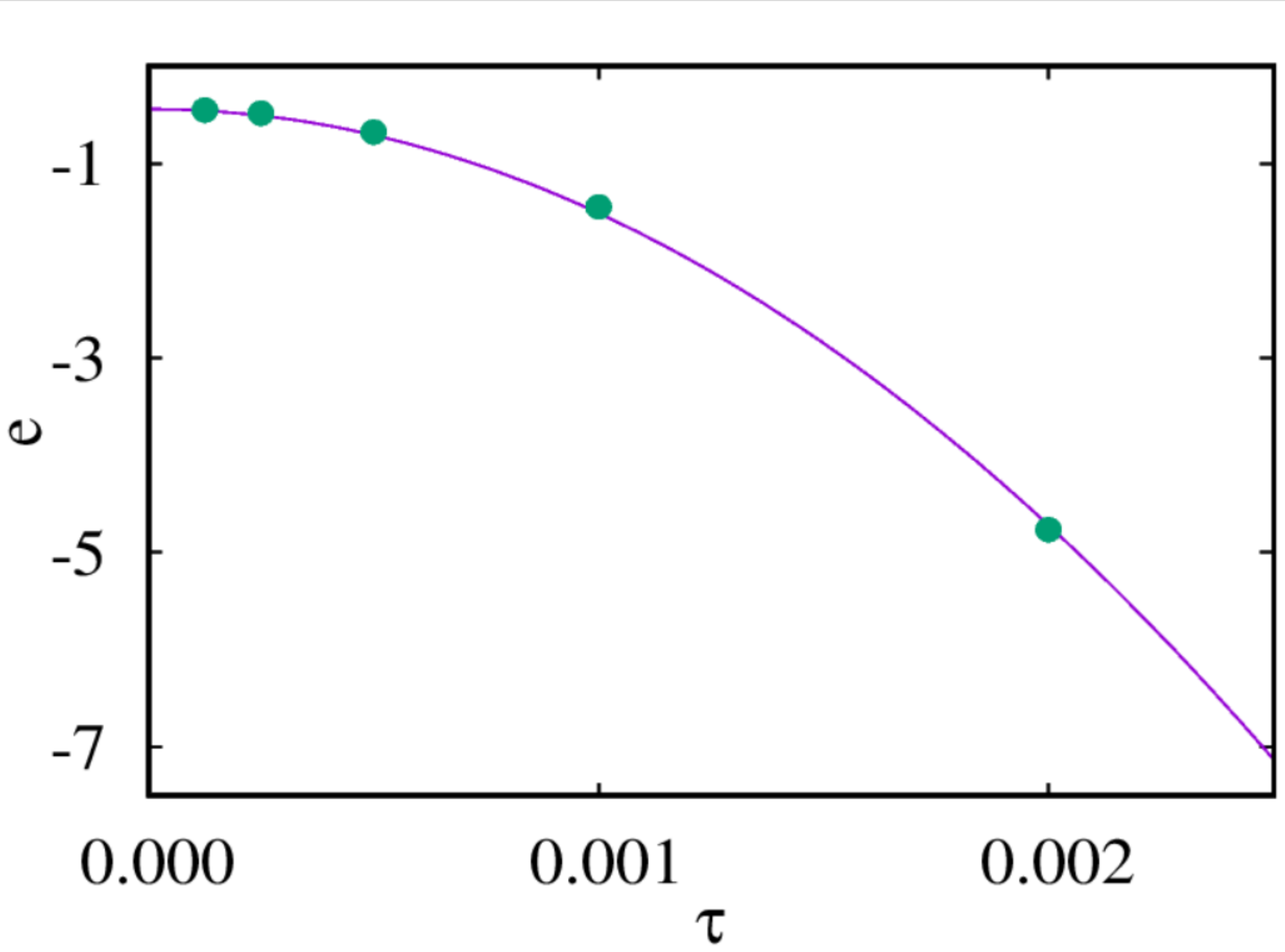}
\caption{Computed energy per particle $e$ for a cluster of $N$=400 dipolar bosons at temperature $T=0.1$, for different values  of the time step $\tau$. The value of $\sigma$ is 0.38. The units are those adopted in this work (see text). Solid line is a quadratic fit to the data. Statistical errors are smaller than symbol size.}
\label{extrapolation}
\end{figure}
\begin{figure}[h]
\centering
\includegraphics[width=\linewidth]{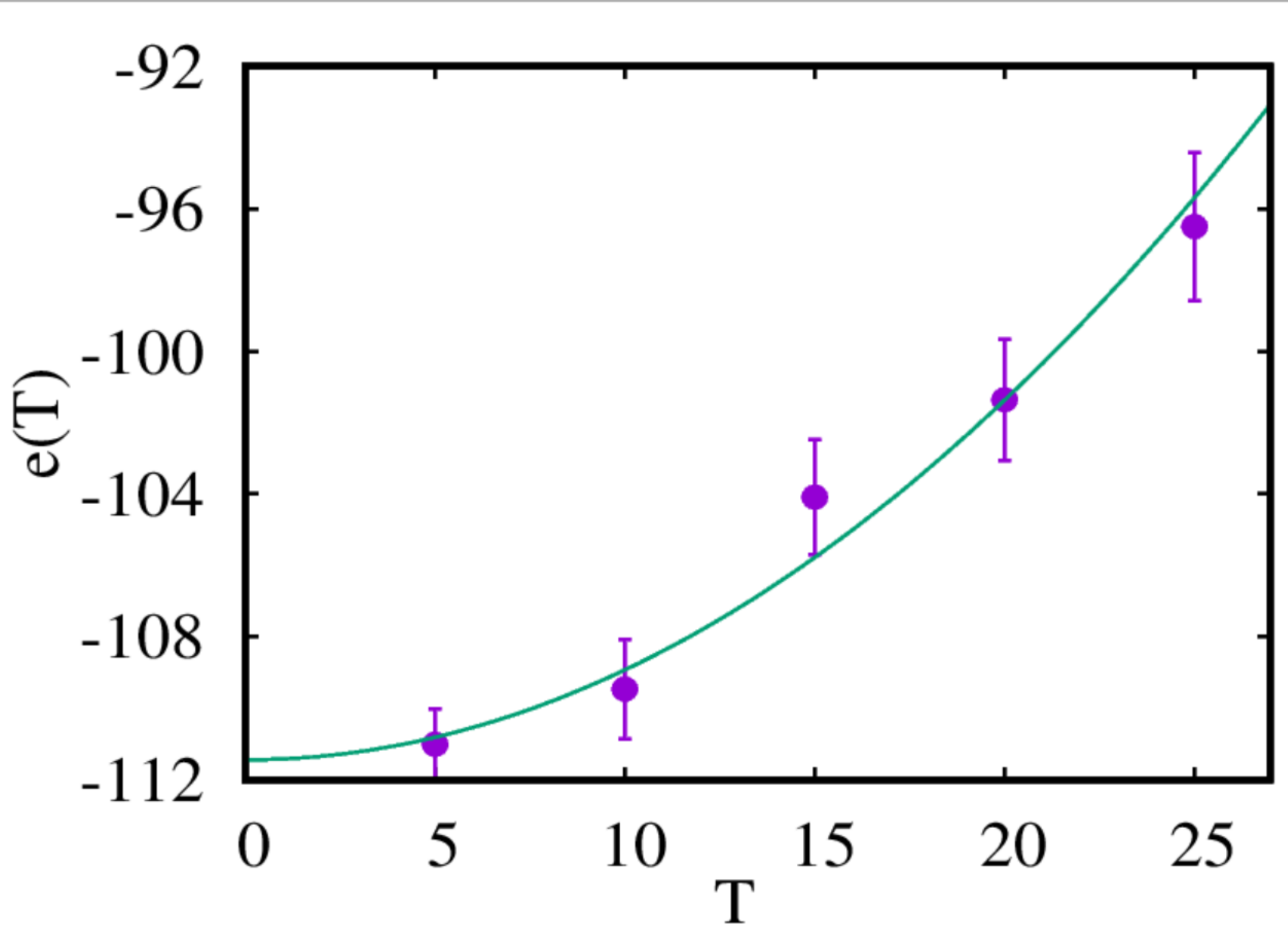}
\caption{Computed energy per particle $e$ for a cluster of $N$=400 dipolar bosons as a function of temperature. The value of $\sigma$ is 0.18. The units are those adopted in this work (see text). Solid line is a quadratic fit to the data.}
\label{T_ext}
\end{figure}

Details of the simulations carried out in this work are standard, and therefore the reader is referred to the original references. As in previous works with dipolar systems, we used here  the so-called ``primitive'' approximation for the short imaginary time ($\tau$)  propagator, which requires a greater number of time slices than more elaborate propagators (e.g., the fourth order one \cite{jltp2}) but is more convenient, due to its simplicity, when dealing with an anisotropic interaction. Obviously, all of the results presented here are extrapolated to the $\tau\to 0$ limit.

Fig. \ref{extrapolation} shows an example of this procedure; specifically, the energy per particle $e$ of the cluster, for a value of $\sigma=0.38$ and a temperature $T=0.1$, is plotted as a function of the time step $\tau$, together with the expected \cite{rmp95} quadratic fit to the data (solid line). The value of $\tau$ for which the physical estimates, within their statistical uncertainties, cannot be distinguished from their $\tau=0$ extrapolation, depends on $\sigma$, ranging from $\sim 5\times  10^{-5}\ \epsilon^{-1}$ for $\sigma=0.18$, to $\sim 5\times  10^{-3}\ \epsilon^{-1}$ for $\sigma=0.49$. As long as a careful time step extrapolation is carried out (which is required in any case, regardless of the short time approximation that one adopts, and which does not prove an excessive computational burden in this case), the use of the primitive approximation causes no loss of numerical accuracy.
\\ \indent
Also dependent on $\sigma$ is the temperature at which the physical estimates are indistinguishable (again, within error bars) from their extrapolated ground state values; at the two extreme of the $\sigma$-range considered here, it is found that the results at $T=10$ ($T=0.02$) can be considered essentially ground state results for the case $\sigma=0.18$ ($\sigma=0.49$). An example of extrapolation to $T=0$ of the energy estimates obtained at finite $T$ is shown in Fig. \ref{T_ext}. Unless otherwise noted, and particularly when comparing results for different values of $\sigma$, extrapolation of the estimates to $T=0$ will always be implied.
\\ \indent
Besides energetic and structural properties, such as integrated density profiles computed along specific directions, we also evaluate the global \cite{skc} and local \cite{paesani,mezz3} superfluid response of the cluster. As we shall see, interesting insight into the physics of the cluster can also be gained by studying the frequency of occurrence of cycles of permutation of different "length", i.e., involving different numbers of identical particles \cite{rmp95}.
\section{Results}
\label{else}
\begin{figure}[h]
\centering
\includegraphics[width=\linewidth]{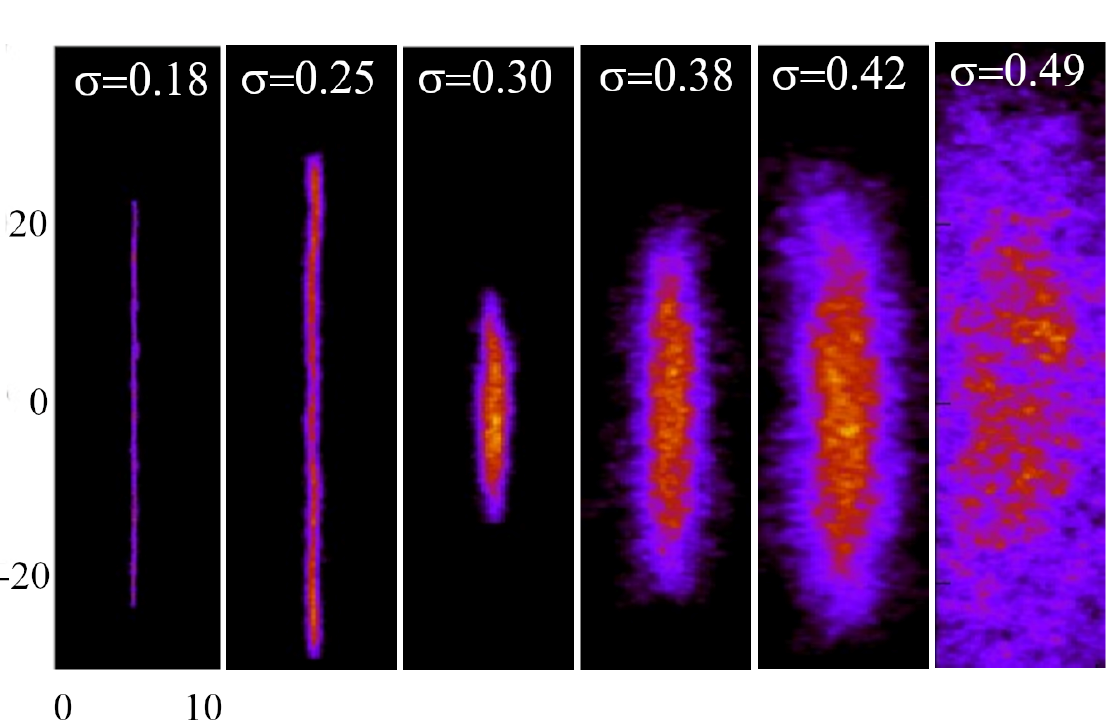}
\caption{Integrated density maps (in the $x-z$ plane) for dipolar clusters of $N=400$ particles at low temperature, for increasing values of the hard core diameter $\sigma$. Brighter spots represent locations of higher density. Lengths are expressed in the units utilized in this work (see text).}
\label{comparison}
\end{figure}
The main finding of this investigation is illustrated in Fig. \ref{comparison}, showing integrated density maps (projected in the $x-z$ plane) computed for the cluster at temperatures sufficiently low that the results can be regarded as ground state  estimates (we come back to this point below). These maps are obtained as averages over a number of representative many-particle configurations collected in the course of sufficiently long computer runs.
\begin{figure}[h]
\centering
\includegraphics[width=\linewidth]{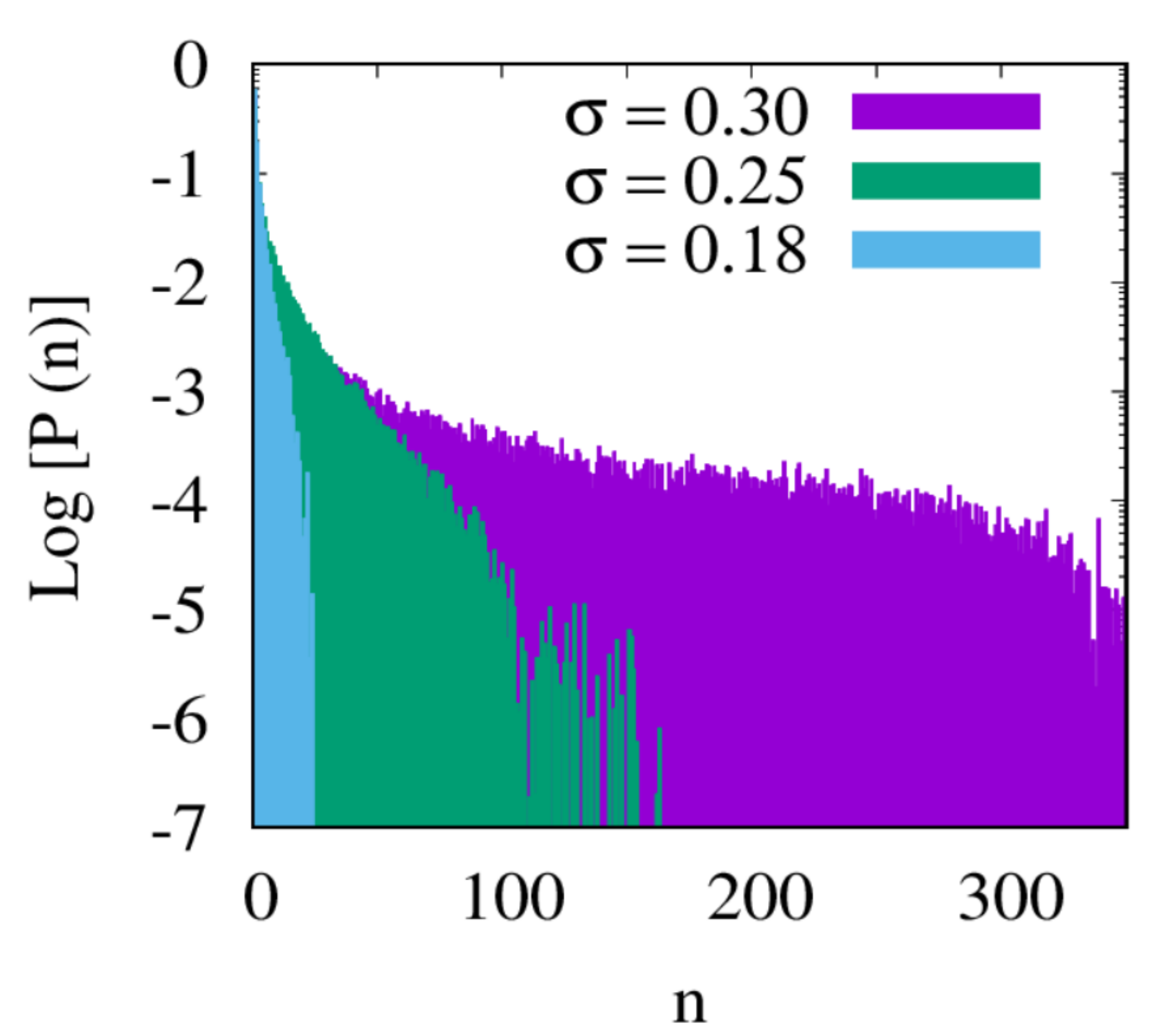}
\caption{Logarithm (base 10) of the frequency of occurrence of exchanges of identical particles in the ground state of a cluster of $N=400$ dipolar particles, plotted as a function of the number of particles involved in the exchange.}
\label{cycles}
\end{figure}
\begin{figure}[h]
\centering
\includegraphics[width=\linewidth]{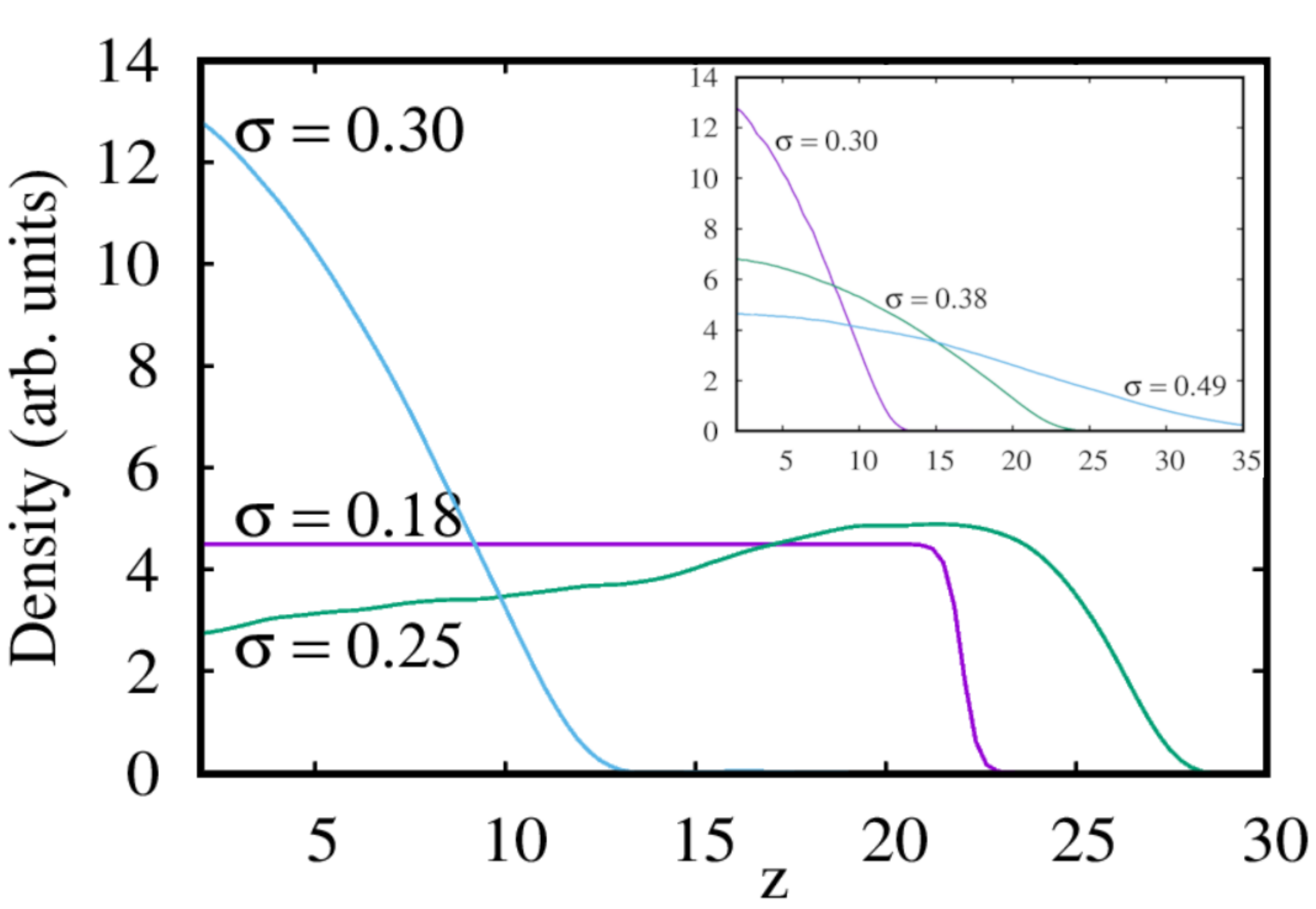}
\caption{Integrated density profiles (along the direction of dipoles alignment, $z$) for dipolar clusters of $N=400$ particles in the $T\to 0$ limit, for increasing values of the hard core diameter $\sigma$, namely 0.18, 0.25 and 0.30. Inset shows the same quantity for $\sigma=0.30$, 0.38 and 0.49. In these plots, $z=0$ is the location of the equatorial plane of the cluster.}
\label{profiles}
\end{figure}
As one can see, for the lowest value of $\sigma$ considered here, namely $\sigma=0.18$ (leftmost panel of Fig. \ref{comparison}, corresponding to a  scattering length $a_s=0.09$) the cluster takes on the form of an essentially one-dimensional filament in the direction of dipole alignment ($z$).
Its length can be very accurately estimated classically, i.e.,  by assuming that the distance between contiguous particles is given by $2^{1/9}\sigma^{4/3}$, i.e., the location of the minimum of the one-dimensional potential $(\sigma/z)^{12}-2/z^3$. As $\sigma$ is increased to 0.25 ($a_s=0.14$), while the cluster remains quasi-one-dimensional in character, nonetheless particle motion in transverse direction is significant; its main effect is that of softening the hard core repulsion at short distance, which in turn results into a filament that, while longer than that with $\sigma=0.18$, is some $\sim 20$\% shorter than predicted classically. Exchanges of identical particles, essentially absent for the cluster with $\sigma=0.18$, begin to be important for this greater value of $\sigma$ (see Fig. \ref{cycles}). The lowering of the density in the central part of the filament, shown by the integrated density profile ($n(z)$) in the $z$-direction (Fig. \ref{profiles}), suggests that exchanges occur primarily in this part of the cluster, as particles tend to keep further away, in order to take part in exchanges while avoiding the repulsive core of the interaction \cite{role}. The weaker axial attraction among particles also results in enhanced zero-point motion, confirmed by the less abrupt, smoother decay of the density profile at the end of the filament, with respect to the $\sigma=0.18$ case (as shown in Fig. \ref{profiles}).

As $\sigma$ is increased above 0.25, the physics of the cluster changes qualitatively, as exchanges of identical particles become considerably more important (see Fig. \ref{cycles}), involving numbers of particles up to $\sim N$ and playing a significant  role in the binding of the system. The most obvious effect is the change of shape of the cluster, from a quasi-one-dimensional filament to a prolate droplet, shrinking in length by a factor three, as assessed from the value of the root mean square distance  of the particles from the equatorial plane ($z_{rms}$) of the cluster (see Table \ref{table}).
\begin{table}[h]
    \centering
    \begin{tabular}{c|c|c|c|c}
       $\sigma$  & $e$ & $z_{rms}$ &$r_{rms}$ &$T$ \\ \hline
       0.18 &$-111(1)$  & $12.8(1)$ &0.14(2) &$5$\\
       0.25 &$-10.7(2)$ &16.0(2) &0.35(2) &1 \\
       0.30 &$-3.1(2)$ &5.73(2) &0.78(2) &0.8\\
       0.38 &$-0.46(1)$ &10.36(4) &1.63(5) &0.1\\
       0.42 &$-0.14(1)$ &12.30(5) &2.32(3) &0.05\\
       0.49 &$-0.02(1)$ &18.0(5) &5.8(1) &0.02\\
    \end{tabular}
    \caption{Energy per particle ($e$) and root mean square particle distance from equatorial plane ($z_{rms}$) and from the axis ($r_{rms}$) of a cluster of $N=400$ dipolar bosons, with different values of the effective hard core diameter $\sigma$ (see text). Also given for completeness is the temperature at which the quantities were computed, which is observed to be sufficiently low to regard the estimates as ground state. The units are those adopted in this work (see text).}
    \label{table}
\end{table}

It is interesting to note that, in previous work \cite{cb} focusing on trapped droplets, each comprising of the order of a hundred particles, a value of $\sigma\sim 0.28$ (i.e., $a_s\sim 0.16$) was identified as that for which a possible ``supersolid'' droplet phase may occur, i.e., one of individually superfluid droplets forming an ordered array, in which global phase coherence may result from long exchanges involving particles in adjacent droplets. 

The results shown here suggest that the occurrence of such a phase reflects the onset of a regime in which droplets switch from being essentially one-dimensional (filaments) to three-dimensional (prolate droplets) objects. This particular value of $\sigma$ at which this occurs, namely $\sim 0.28$, can be identified as that for which the potential energy of dipolar attraction, which favors particle alignment in the direction of the dipole moment, is quantitatively comparable to the reduction of kinetic energy  arising from quantum-mechanical exchanges (i.e., the effect underlying Bose condensation in the bulk), which result in the cluster taking on a less markedly one-dimensional, more cigar-like shape.
    
On further increasing $\sigma$, the cluster becomes progressively less bound;
as the binding energy approaches zero, and its precise determination becomes increasingly difficult, an additional empiric criterion to establish whether the cluster is self-bound in the $T\to0$ limit is the observation of significant particle evaporation. The largest value of $\sigma$ for which a bound state appears to exist at low $T$, is $\sigma=0.49$, i.e., clusters with $N=400$ particles are unbound for $\sigma\gtrsim 0.5$, corresponding to a scattering length $a_s=0.31\ a_D$. 

Two observations can  be made. The first is that, as shown by the results in Table \ref{table}, as $\sigma$ increases the ratio $z_{rms}/r_{rms}$ of the root-mean-square particle distance from the equatorial place to that from the axis of cluster (in the direction of dipole alignment) decreases, but remains as large as $\sim 3$  as the unbound limit is approached, i.e., a self-bound  droplet retains an elongated shape. The second observation is that, for a fixed $N$, the length of the cluster in the axial direction depends non-monotonically on $\sigma$, reaching comparable maximum values in both the classical and quantum regimes.

All the clusters with $\sigma=0.30$ and higher are 100\% superfluid at low temperature, and the superfluid response is uniform throughout the cluster, within the precision of the calculation. 
As mentioned above, clusters with $\sigma \lesssim 0.3$ are largely one-dimensional, and therefore the attribution of superfluid properties becomes ambiguous, given the vanishing moment of inertia with respect to the relevant axis. 
    
\section{Discussion}
\label{conc}
We have carried out extensive numerical simulations of a self-bound, free standing cluster of few hundred dipolar Bose particles with dipole moment all aligned. The system is modeled in the same way as in other published simulation work, namely including a short-range repulsion at short distance in order to prevent the system from collapsing in the presence of the sole three-dimensional dipolar interaction. We study the physical properties of the cluster as a function of the effective hard core diameter $\sigma$  of the short-range repulsive potential.

We find that two different physical regimes can be identified, one that corresponds to values of $\sigma \lesssim 0.28\ a_D$, $a_D$ being the characteristic length of the dipolar interaction, wherein the cluster is essentially a one-dimensional filaments, in which particle motion is confined to the axis of the filament and exchanges of indistinguishable particles are rare. Binding comes almost entirely from the attractive part of the dipolar interaction and the behavior of the system can be largely understood along classical lines.

For $0.28\ a_D \lesssim \sigma\lesssim 0.50\ a_D$ the cluster remains self-bound but its shape is that of a prolate droplet, and exchanges of indistinguishable particles play a significant role in keeping the cluster together. In this range, clusters are superfluid at low $T$, and much of their physics is reminiscent of that of $^4$He clusters, of course with the important difference that self-bound dipolar clusters are non-spherical.

We now discuss the possible relevance of these calculations to current  experiments \cite{pfau}, in which typical number of particles is one to two orders of magnitude greater than that considered here. 
We begin by noting that the choice of number of particles made in this study, namely $N=400$, is that which allows for the observation, in a numerical simulation, of stable self-bound clusters in a sufficiently wide range of $\sigma$, in which clusters display both classical and quantum behavior. Clusters of significantly smaller size (e.g., $N=100$) remain self-bound only in the ``classical'' $\sigma\to 0$ limit.

Experience accumulated over decades of theoretical studies of clusters of helium and parahydrogen suggests that the most important physical properties (e.g., aspect ratio, binding energy, superfluidity) of a {\em self-bound, free standing} cluster of few hundred particles are not significantly different from those of a cluster with one or two orders of magnitude more particle. For example, the binding threshold found here for a cluster of $N=400$ particles, namely $\sigma\sim0.5$, is likely to be very close to the saturation (bulk) value, for which no theoretical estimate has yet been obtained, as the threshold to binding in one dimension for the bulk phase has been determined \cite{kora1d} to be $\sigma\sim 0.65$, and binding is generally weakened in higher dimensions.

In the quantum regime, where the interaction is dominated by the repulsion at short distance and the cluster becomes nearly spherical as $\sigma$ increases, there seems to be no physical reason not to expect a simple uniform expansion of the cluster as $N$ grows, proportional to $N^{1/3}$ in every direction. Thus, for example, the axial root-mean-square length of a self-bound droplet of $N=40,000$ particles can be expected to be $\lesssim 100\ a_D$, using the values of Table \ref{table}.

In all present experiments clusters are harmonically confined in all directions. In order to assess the importance of confinement on the properties of a dipolar cluster, one may estimate the characteristic confinement energy, in the units adopted in this work. This can be roughly expressed as $z_{rms}^2/2\xi^4$ in the axial direction (from which the most important contribution comes) and $\xi$ the harmonic confinement length. For a typical value of $\xi \sim 40\ a_D$ and taking as reference the results given in Table \ref{table}, one arrives at a harmonic confinement energy  $\lesssim 10^{-3}\ \epsilon$, i.e., 
negligible except for very weakly bound clusters. 

Thus, the results obtained in this work show that typical experiments with droplets comprising up to at least $\sim 10^4$ particles, and for values of the scattering length $a_s \lesssim 0.3\ a_D$, can be expected to probe the physics of self-bound droplets, the confining trap not playing a significant role.

\section{Declaration of Competing Interests}
The author declares that he has no known competing financial interests or personal relationships that could have appeared to influence the work reported in this paper.
\section{Acknowledgments}
This work was supported by the Natural Sciences and Engineering Research Council of Canada.

 \bibliographystyle{elsarticle-num} 
 \bibliography{refs}





\end{document}